*Review Article*

# Use of Physicochemical Modification Methods for Producing Traditional and Nanomodified Polymeric Composites with Improved Operational Properties

**Aleksandr E. Kolosov** [ID],[1] **Volodymyr I. Sivetskii,**[1] **Elena P. Kolosova,**[1] **Volodymyr V. Vanin,**[1] **Aleksandr V. Gondlyakh,**[1] **Dmytro E. Sidorov,**[1] **Igor I. Ivitskiy,**[1] **and Volodymyr P. Symoniuk**[2]

[1]*National Technical University of Ukraine "Igor Sikorsky Kyiv Polytechnic Institute", 37 Prospect Peremohy, 03056 Kyiv, Ukraine*
[2]*Lutsk National Technical University, 56 Potebny Street, 43018 Lutsk, Ukraine*

Correspondence should be addressed to Aleksandr E. Kolosov; a-kolosov@ukr.net

Various aspects of the methods of physical and physicochemical modification of components of filled thermoplastic composite materials are analyzed, aimed at improving the surface properties of the fillers and the technological properties of the polymer matrix during their interaction. It is noted that the improvement of the interfacial interaction of the components of polymer reactoplastic composites, including adhesive strength, is a key factor for improving the reliability of the cured filled composite. As a promising area of research, a modification of the surface of the reinforcing fibrous filler and the technological characteristics of the liquid polymer binder, aimed at increasing their contact properties in the composite, was chosen. The effectiveness of the physical method of modifying the components of composites in the form of low-frequency ultrasonic processing is described. The peculiarities of cluster formation and physicochemical modification of epoxy polymers filled with dispersed fillers are analyzed. Attention is focused on the effectiveness of ultrasonic processing in the cavitation mode for deagglomeration and uniform distribution of nanoparticles in a liquid medium during the creation of nanocomposites. Experimentally confirmed is the improvement of the technological properties of liquid epoxy polymers, modified by ultrasound, used for the impregnation of oriented fibrous fillers, as well as the improvement of the physicomechanical properties of the sonicated epoxy matrices. Some issues of biological modifications of components of polymers for functional application are briefly reviewed.

## 1. Introduction

At the present stage of development of science and technology, polymer composite materials (PCMs) are a fundamental element in the implementation of innovative solutions involving the creation of new designs and devices, as well as appropriate technologies. These PCMs are widely used in the aerospace industry, aircraft, sports, biology, chemistry, medicine, electric power industry, microelectronics, machine building, military [1], and other industries.

The study of the physicochemical aspects concerning the surface properties of the constituent components of PCMs is the foundation of the subsequent wide practical use of the results of such research in the form of creating innovative technologies. In particular, improving the interfacial interaction of the constituent components of the PCMs is a key factor for improving the reliability of the cured filled composites [2].

To do this, in practice, apply a range of methods of modification, including physical, chemical, and combined physicochemical modification of the components of the filled composites. Moreover, the main components of the composite are the filler and the liquid polymer matrix (polymer binder (PB)). In this section, an important role belongs to the study of adhesion, namely, adhesive strength, depending on the conditions of molding PCMs, in particular, the method of modification used.

It is known that adhesive strength is a multifactorial indicator, depending primarily on the nature of the liquid polymer and substrate, as well as on the conditions for the formation of a structural compound, including polymer coating. Coverings of monomeric and oligomeric film-forming agents, which are transformed into a polymer (three-dimensional) state directly on the substrate, have the highest adhesive strength.

In some cases, monomers and oligomers are capable of chemisorbing on the surface of metal substrates. Subsequent polymerization or polycondensation of monomers and oligomers leads to the formation of grafted polymers that are chemically bound to the substrate, for example, in the form of a nonmetallic or metallic surface. In the case of reactoplastic polymers, there is often a correlation between the adhesion strength and the cohesion of the film coating material.

One of the effective ways to increase the adhesion strength of a joint while simultaneously increasing the productivity of the molding process is the physical modification method, which consists in using a contact ultrasonic (US) field [3, 4]. Moreover, US treatment can be subjected to both the initial composition (liquid or even powder) before they are applied to the substrate surface and the polymer coatings immediately at the time of their formation [5].

In the latter case, by varying the basic processing parameters, in particular, the irradiation dose, time, and intensity of the US effect, you can obtain coatings (epoxy, epoxy-furan, polyethylene, etc.) with adhesive strength, which in some cases are times the strength of the corresponding untreated coatings. Therefore, research in the field of design and projecting on the basis of the structural-parametric method of both existing PCMs and new PCMs with improved surface properties is the current direction of polymer material science and polymer technology [6].

A promising area of research in polymer technology is the modification of the surface of reinforcing fiber filler (FF) and liquid PB to increase their contact properties in the composite. The optimal modification of the above components leads to the improvement of the physicomechanical and operational properties of the composites, moreover, both traditional PCMs and nanomodified (NM) PCMs [7]. For example, carbon fiber composite materials with improved performance properties are a typical example of the implementation of the modification of the surface of carbon fiber and PB. At the same time, ultradisperse carbon nanoparticles are uniformly distributed in the volume of a liquid PB, while macrofiber carbon plastics, in turn, are surrounded by this combined PB [8, 9].

Improving the performance properties of NM PCMs is also directly related to the improvement of the interfacial interaction of their constituent components. Depending on the composition and properties of the components, including the magnitude of the surface energy, as well as the features of the composite manufacturing technology, it is possible to obtain NM PCMs of various functional purposes [10, 11]. At the same time, in addition to analyzing promising areas for the development of NM PCM technologies, economic factors for the implementation of such technologies are important, which predetermine the feasibility of industrial implementation of completed projects [12].

It is well known that the main difficulties in obtaining NM PCMs are due to the need for uniform distribution of nanoparticles in a liquid oligomer (polymer) to ensure maximum contact surface between the liquid polymer system and nanoparticles incorporated into it, especially based on carbon [13]. In this case, US treatment in the cavitation mode is probably the most effective means for achieving the deagglomeration of nanoparticles in a liquid medium [14].

In addition, effective US impact contributes to achieving a virtually uniform distribution of deagglomerated carbon nanofiller particles in the liquid polymer matrix [15]. This, in turn, is the most important factor contributing to the maximum realization of the individual properties of the nanofiller in the composite, including due to the increase in the total effective contact surface area of the particles due to their environment of PB [16].

When producing NM PCMs, it is important to establish (as a rule, experimentally) an effective "homeopathic" dose (range of dose values) of nanoparticles incorporated into a liquid polymer, since the operational (functional) properties of the final nanocomposite depend on it. Moreover, the deviation of the content of nanofiller in one direction or another from the boundaries of the effective range, as a rule, leads to the deterioration of the final properties of the nanocomposite. No less important are the parameters of the technological process of homogenization of nanoparticles incorporated into a liquid medium (time, pressure, temperature, and other specific process parameters inherent in a particular molding technology) when creating designs from NM PCM of deterministic functionality [17].

We should also mention another class of innovative PCMs, namely, intellectual (smart) PCMs [18–20], including NM PCMs [21]. PCM intellectualization is achieved by modifying polymer composites with components that contribute to the transformation of such materials into materials that self-diagnose and adapt to external temperature-force effects.

This direction of research is extremely important and, apparently, will determine the future development of innovative polymer technology.

The possible directions for improving the technological properties of traditional and NM PCM according to the considered approach are shown in Figure 1.

The purpose of the article is a brief analytical review of some existing and developed by the authors methodological approaches to create both traditional PCMs and NM PCMs with improved surface properties through the selection and subsequent physical and physicochemical modification of the components that make up the composite according to the mentioned approach (see Figure 1).

The following directions can be identified:

(i) Modification of liquid polymer compositions on the example of epoxides

(ii) Cluster formation and physicochemical modification of epoxy polymers (EPs) with dispersed fillers

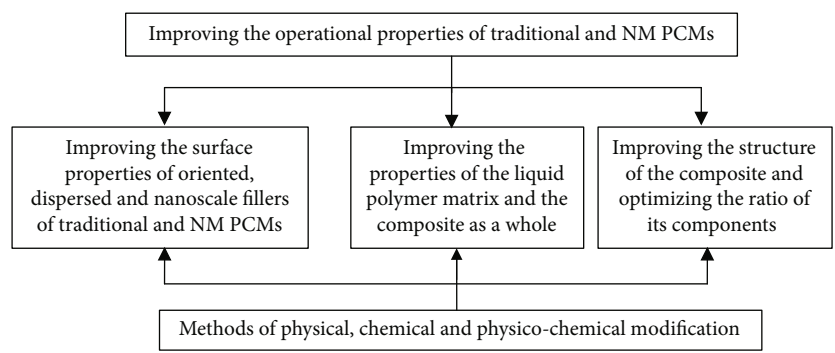

FIGURE 1: The considered approach to improving the technological properties of composites.

(iii) US modification of the components of the polymer composite

(iv) Physical modification of the FF surface

(v) Technological properties of liquid EP, modified by US

(vi) Physical and mechanical properties of sonicated epoxy matrices

(vii) US modification upon receipt of epoxy nanocomposites

(viii) Some methods of physicochemical and biological modifications of components of polymers of functional application

The above aspects are shortly described in this review article.

## 2. Results and Discussion

*2.1. Modification of Liquid Polymer Compositions on the Example of Epoxides.* It should be noted that the modification of the components of existing structural PCMs, used for the manufacture of a wide range of products, is no less important than the creation of new PCMs [22]. This modification is used for deterministic regulation of the structure and properties of the resulting polymer composite at different levels of the technological process of its manufacture. At the same time, among the PCMs occupy an important place EPs.

Therefore, an important task is to improve the technological and operational characteristics of epoxy PCMs. This complex task includes increasing the viability, reducing the technological viscosity, improving the deformation and strength properties, heat, and bio- and chemical stability, increasing the dielectric properties, reducing flammability, and improving the economic efficiency of producing such composites [23]. Among the modern methods of polymer modification, there are three main methods of modification, namely, chemical, physical, and combined physicochemical.

The purpose of this work is a more detailed consideration of the methods of physical and combined physicochemical modification.

*2.2. Methods of Physical Modification of Liquid Epoxy Compositions.* Methods of physical modification of liquid ECs include, for example, the following: heat treatment, radiation exposure, vacuum compressor treatment, periodic deformation, and processing of polymers in magnetic and electric fields [24].

Physical modification of liquid polymers can be carried out at different stages of production, processing, and subsequent use. Often, physical modification is used in combination with chemical or physicochemical methods of modifying a liquid polymer.

Heat treatment is one of the most common and effective methods to improve the performance properties of solidified EPs. This treatment affects the molecular mobility, structural order, speed, and depth of curing of the polymer, depending on the temperature-time curing. As a result of optimal heat treatment, an improvement in the physical properties of EPs is observed, in particular, hardening (on average by about 10–15%).

The main technological parameters of the heat treatment of the liquid polymer are the temperature and duration of curing, as well as the rate of heating and cooling. In turn, the choice of the curing temperature of a liquid EO depends on its type and the hardener used.

As a rule, EPs that harden at elevated temperatures (i.e., EPs obtained on the basis of "hot" curing ECs) are characterized by increased strength and rigidity (on average by about 10–20%). This result is explained by a change in the supramolecular structure of the epoxy oligomer (EO) [25].

The appearance and size of polymerized supramolecular formations in a polymer also depend on the solidification temperature. At the same time, as a result of an increase in the hardening temperature, the size of the associates decreases, as well as their lifetime, which is explained by the intensification of the thermal motion of molecules under the influence of the temperature factor. An increase in temperature also affects the reduction of the size of the formed globules, the size of which, in turn, depends on the density of the molecular network, which, in turn, depends on the parameters of the elastic properties of EPs.

In this case, polymers with a thin spherical structure, as a rule, are more durable compared to polymers with a thickened and nonspherical structure. This is due to the fact that a decrease in the size of the globules due to the temperature factor leads to an increase in the probability of chemical and physical interactions of molecules, which is ensured by the functional groups located on the surface. In turn, an increase in the cure depth leads to a shift in the peak of the glass transition temperature of reactoplastic polymers towards higher temperatures. This leads to a reduction in curing time and to a corresponding acceleration of the formation of EP.

Another method of temperature modification is associated with heat treatment of already hardened polymers. Using this method leads to a number of positive results. Among them, the reduction in the number of functional groups of EO and hardener, which did not react, due to the preliminary curing of the polymer, and the increase in the depth of curing of the polymer are particularly significant. The consequence of this is to increase the performance characteristics of the EP (up to 8%).

One of the most effective methods for improving the technological and operational characteristics of liquid polymer media is vibration or low-frequency US treatment [26, 27]. Under the influence of the above vibroacoustic effects on the effective modes, the viscosity of the liquid polymer compositions decreases sharply (up to 30–50%) due to the "injection" of energy into the system and its heating. The conditions for homogenizing and postprocessing these compositions are also improved.

Due to the effective vibration effect on the liquid oligomer, before its hardening, a less defective and more ordered polymer structure is observed, which results in hardening of the vibroprocessed polymers. At the same time, it is necessary to take into account the slight relaxation time of the vibroprocessed composition, which is the limiting technological factor of the vibroprocessing. Due to the effective modes of US treatment, the conditions for homogenization of the composition and the kinetics of its curing are improved, and the dynamic viscosity of the composition decreases (on average by about 20–40%), and its relaxation time increases (compared to the vibroprocessed composition) on average 1.5–2 times.

Also, as a method of physical modification, processing of polymers in a magnetic field is used. Such processing at effective modes leads to a decrease (on average by about 15–25%) in viscosity and an increase (on average by about 10–20%) in the homogeneity of the polymer mixture, the formation of a more ordered structure of the hardened compositions. The consequence of the above factors is the improvement of the operational and physicomechanical characteristics of EPs processed in a magnetic field (on average by about 10–20%).

Radiation hardening of epoxies at optimum gives similar results. In addition to the above methods of physical modification, the methods of modification by low-temperature and electrothermal treatment (high-frequency currents), ultraviolet and infrared radiation, etc., are widely used.

In practice, to give the created polymer materials special (functional) properties, a combination of the above several methods of physical modification is used.

### 2.3. Physicochemical Modification of Liquid Epoxy Compositions.
Controlling the properties of epoxy resin by incorporating various components is one of the most widely used physicochemical methods for modifying the EP structure. Such a physicochemical modification allows the pro-duction of materials, including functional, with the necessary physicomechanical and operational properties [28]. Physical and chemical modification is carried out, as a rule, by incorporating solid insoluble, microdispersed fillers, inert plasticizers, solvents, stabilizers, functional dopants, surfactants, and a number of other modifiers into the matrix of the liquid polymer before its polymerization.

At the same time, the filling is the simplest, widely used, and highly efficient method of directional control of the physical, mechanical, and operational properties of EPs. Optimal filling leads to an increase in mechanical strength and rigidity, chemical and heat resistance, dielectric properties, etc.

In the general case, physicochemical modification of polymers in the form of filling is understood as the combination of liquid polymers with solid and (or) gaseous substances. In this case, gaseous substances are relatively evenly distributed in the volume of the liquid polymer composition and have a clearly defined interface with a continuous liquid polymer phase (matrix).

In practice, solid finely dispersed fillers are mainly used for the preparation of filled liquid polymer compositions. Among such fillers, most often used are fillers in the form of particles of spherical shape (glass or ceramic microspheres, fly ash), granular form (soot, silica, wood flour, chalk, and kaolin), lamellar form (talc, graphite, and mica), and needle-shaped (oxides, salts, and silicates), as well as fibrous fillers (FF) (cotton, glass, organic and carbon fiber, asbestos, and cellulose).

Depending on the degree of filling with fine fillers, the functional purpose of the PCMs varies. For example, traditional reactoplastic PCMs with a low degree of filling (the relative degree of filling $v$ ranges from 0 to 0.3) have high values of deformation (at least 10–15%) and viscosity (at least 20–30%), but low static strength and rigidity (approximately 10–15%). At the same time, highly filled structural PCMs (the relative degree of filling $v$ is from 0.3 to 0.7) have higher values of rigidity and compressive strength (at least 15–20%) with respect to unfilled polymers. Such highly filled composites have high (approximately 10–15%) brittleness and a low (approximately 2–5%) deformation limit, which localizes the spheres of their use.

Aspects of physicochemical modification of EP with dispersed fillers are described in more detail below.

### 2.4. Clustering and Physicochemical Modification of Epoxy Polymers with Disperse Fillers.
Physicochemical modification of EPs with dispersed FF has its own characteristics. In the general case, such a physicochemical modification is usually considered from the standpoint of cluster analysis, features of surface interaction of fillers with liquid EO, manifestations of the mechanism of molecular interaction of EP and filler, and the theory of adhesion between EP and filler. This article briefly discusses aspects of physicochemical modification only from the standpoint of cluster analysis and also analyzes the feasibility of using US to increase the effectiveness of such a modification.

The features of cluster formation in a filled liquid polymer composite are as follows. A cluster is a group of particles of a dispersed filler, which are separated by thin polymer layers and which are completely in the film phase. As a rule, cluster structures are formed from dispersed particles as a result of a series of processes (diffuse, sedimentary, etc.) that are associated with the involuntary relative movement of particles of the dispersed filler in the liquid polymer.

For example, cluster structures are formed as a result of the forced movement of particles of dispersed filler in the process of homogenization of the liquid polymer matrix. The formation of clusters begins with the contact interaction of two separate dispersed particles. Due to the fact that the structure of the boundary polymer layer is formed as a result of the tendency of these dispersed filler particles to reduce their surface energy, it is energetically more advantageous when the limiting (facel) layers of individually dispersed particles begin to interact with each other in contact.

The result is an uneven distribution of dispersed particles, which, however, helps to compensate for the excess surface energy. At the same time, particles of the dispersed filler begin to be structured in such a way that the liquid polymer in the space between the dispersed particles completely passes into an ordered state with the formation of mainly linear clusters. Thus, cluster *1* is formed, in which dispersed particles with a conditional radius $r_c$ are arranged along a curve or some conditional curved line (see Figure 2).

At the same time, it should be noted that the increase in the length of linear clusters does not occur infinitely, but only to a certain specific size. This limiting length is determined by the state in which the cluster becomes hydrodynamically unstable, as a result of which it splits into several small linear clusters, or new ring cluster *2* is formed (see Figure 2). The formed ring cluster *2* is grouped together, forming after grouping spatial cell clusters in the form of irregular (distorted) spheres.

The peripheral layer of the formed spatial cell cluster consists of particles of dispersed filler, which alternate with the film phase of the polymer matrix. The inner region of the sphere of a spatial cell cluster no longer contains particles of dispersed filler but contains only a liquid polymer matrix in the bulk state. At the same time, if the used filler is polydisperse, when it is combined in the liquid polymer, small particles are "captured" by the surface of larger particles, as a result of which dense pseudospherical accumulations of particles appear.

It should also be noted that the clusters, which are freely distributed in the volume of the liquid polymer material and are not connected together (i.e., do not contact), do not have a reinforcing effect on the hardened polymer. At the same time, in case of destruction of a polymer filled with a dispersed filler, they can only serve as a stopper for emerging cracks, reducing the speed of their propagation in the polymer under the action of loads.

That is, the presence of unbound clusters leads to an increase in the crack resistance of solidified composites filled with dispersed fillers. In this case, the hardening of the hardened composite filled with dispersed particles occurs only when spatial frameworks of dispersed filler particles and the phase of a polymer matrix film are formed in the volume of this composite. Moreover, the transition from local clusters to the skeleton of a bulk polymer matrix occurs as a result of the unification and enlargement of these local clusters.

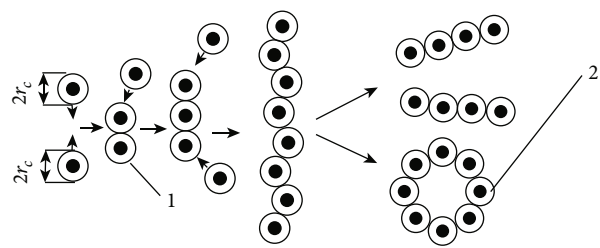

FIGURE 2: The conventional scheme of the formation and subsequent transformation of clusters in a liquid polymer composite: *1*: elementary two-particle linear cluster; *2*: transformed ring cluster.

At a certain stage of filling, the entire volume of the composite filled with dispersed filler is penetrated by one infinite cluster with the formation of a spatial skeleton. The last transformation in clustering leads to the hardening of the composite. Considering the above stages of cluster formation, it can be reasonably assumed that US processing of a liquid polymer matrix in the cavitation mode, as well as contact modification of the surface of dispersed excipients with US waves to increase (maximum "release") their contact surface area and uniform distribution in the polymer, can be considered as a reinforcing clustering factor.

In addition, the dispersion (characteristic size, for example, the nominal diameter of the sphere) of the filler particles also has a significant impact on the strength of the filled PCM. This is determined by the value of the specific surface area of the filler per one dispersed particle. Thus, an increase in the particle size of the dispersed filler leads to an increase in their surface area, that is, to a decrease in their surface energy and, therefore, cohesive bond energy, which leads to a decrease in the strength of the filled composite. The reverse is also true: with a decrease in the dispersion of the filler (and with its uniform distribution in the liquid polymer), the strength of the dispersed PCM increases.

### 2.5. Ultrasonic Modification of Polymer Composite Components

#### 2.5.1. Physical Modification of the Surface of Fibrous Filler. To improve the quality of the impregnation of capillary-porous FFs with liquid PBs, besides a number of other factors, it is advisable to carry out surface modification of the used FFs. This can be implemented, for example, in the form of activation of the FF surface by contacting them with US in the presence of a liquid EC layer. A study of the contact wetting angles of a liquid PB surface of such FF (Figure 3) shows that on the rougher surface FF (Figure 3(a)), the wetting angles $\theta$ are higher and vice versa.

From this fact, it was concluded that the reduction of the surface roughness of the FF contributes to a more rapid impregnation of its structure with liquid PB, which can be explained as follows. A drop of the impregnating solution applied to the rough surface of the FF is located first on the outer surface of the protruding fibers, as shown in

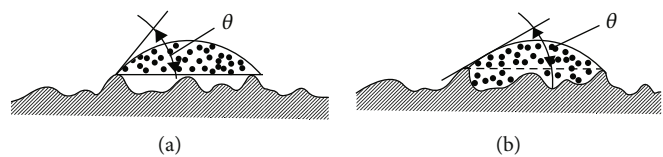

FIGURE 3: Schematic representation of the behavior of a drop of an impregnating solution PB on a rough surface of FF: (a) primary stage; (b) a drop of the impregnating solution after spreading over the FF surface.

Figure 3(a). This is due to the fact that the adhesion force of molecules of liquid PB (cohesive strength) is stronger than their adhesion to the surface of the fibers.

Then, after the destruction of these adhesion forces, for example, by US activation, the surface of a drop of liquid PB can spread and fully come into contact with the surface of the FF. In this case, liquid PB spreads, fills the air gaps that were under the drop of PB, and then freely penetrates deep into the capillary FF (Figure 3(b)). At the same time, there is a sharp decrease in the wetting angle $\theta$, which characterizes the degree of propagation of liquid PB on the FF surface, or penetration into the depth of its capillaries.

From the above, it was concluded that a higher initial wetting angle $\theta$, if it decreases sharply in time, indicates a good impregnation of the capillary-porous structure of the FF [29]. Effective wetting by liquid PB of the FF surface in the process of "wet" winding [30] with optimal operating parameters, in particular, the tension force FF during its winding [31], improves the structure of the hardened unidirectionally reinforced composite, which is confirmed by its qualitative and quantitative results of its microstructural analysis of cross section [32, 33]. In particular, as a result of effective chemical modification of the surface of the macrofiller and an increase in its wetting ability $\sigma \cdot \cos \theta$, where $\sigma$ is the surface tension coefficient, the porosity of the structure of the hardened traditional epoxy PCM can decrease on average from 7-10% to 3-4%, depending on the type of macrofiller used and the polymer matrix.

This is also confirmed by the comparative results of experimental and theoretical histograms of the distribution of the diameters of unidirectional fibers and the distances between them, constructed for hardened composites, which were formed with different regime parameters. At the same time, the integral characteristics, namely, the effective capillary radius (half of the average distance between the fibers), porosity, and the specific surface of the unidirectional reinforcement as a capillary-porous body impregnated with a liquid PB, were selected as the structural parameters of unidirectionally reinforced traditional PCMs.

Regarding the qualitative assessment of the adhesive properties of fibrous macro- and microfillers, it should be noted that, unlike glass fibers, another common type of fillers for reactoplastic PCMs, namely, carbon fibers (CFs), has a lower (approximately 2–2.5 times) wettability [34]. One of the obvious solutions to improve the wettability is chemical modification of the surface of the CF, for example, by applying a certain catalyst to their surface by the method of "free" capillary impregnation or by another method (in particular, modification of the surface of CNTs). However, there may be difficulties associated with the insufficient wettability of the CF surface by liquid PB (see Figure 4).

Therefore, when using the deposition of a salt of a catalyst from a solution in the process of chemical modification of the surface of CF, it is necessary to preliminarily prepare its surface to improve its wettability. This can be done in several ways, for example, by functionalizing the surface, by increasing the specific surface area, or by developing a porous structure of the CF.

## 2.6. Technological Properties of Liquid Epoxy Polymers Modified by Ultrasound.

The effectiveness of the implementation of the process of US modification involves determining the optimal parameters of such processing, in particular, the intensity $I$ and the frequency $f$ of US vibrations [35, 36]. At the same time, US cavitation is the dominant factor for sonicated liquid media, and US impregnation of fibrous fillers has a sound-capillary effect [37, 38]. This allows you to determine the modes and effects of US treatment used in relation to specific liquid media [39, 40].

So, for example, the following operational parameters were used to sonicate the EO: frequency $f = 17 - 44$ kHz, amplitude $50 - 120 \mu m$, intensity $I = 15 - 30$ W/cm$^2$, temperature $T = 70 - 90°C$, time $\tau = 30 - 45$ min, and bulk density of the input energy $<\omega> = 1.2 - 2.5$ W/cm$^3$. For the PB was used EO brand ED-20 (100 wt.p.), which was cured with diethylenetriamine (DETA) in an amount of 10 wt.p.. As for the technological characteristics of the EO, the minimum equilibrium contact angle (wetting) of the EO on the glass substrate $\Theta$ min at a temperature of 20°C, the height of the impregnation $h$ EO, and the wetting ability of $\sigma \cdot \cos \Theta$ on fiberglass were investigated (see Figure 5).

An antibate dependence of the wetting angle $\Theta$, the wetting ability of $\sigma \cdot \cos \Theta$, and the height of impregnation $h$ for low frequencies of US $f$ depending on the sonicating time $\tau$ was experimentally established. In this case, the effective time range of sonicating time $\tau$ lies within 25-35 minutes for this type of oligomer ED-20, which predetermines the optimal duration of sonication in the studied modes.

It is also necessary to take into account the frequency range of sonication of liquid media, for example, low-frequency and midfrequency ranges $f = 1 - 1.5$ MHz [41]. In this case, the sonication in the specified ranges can occur both separately and simultaneously (in the latter case, when a combination of the specified frequency ranges). Another important parameter of US treatment of liquid media is static pressure $P = 0.4 - 0.6$ MPa [42], the appropriateness of which is dictated by the need to increase (approximately 15–20%) the intensity $I$ and reduce (approximately 20–25%) the time of sonication $\tau$.

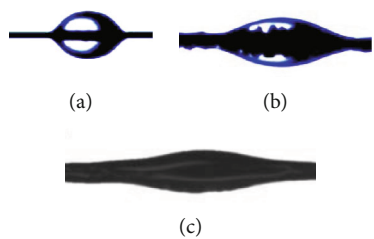

FIGURE 4: Behavior of a drop of epoxy resin on the filament of the original (a) and modified CNT (b, c) CFs: (a) original fiber; (b, c) CF, modified with CNT.

An increase in the maximum lift height $h$ of the PB along the filler fiber during US treatment indicates the possibility of intensifying the capillary impregnation process [43]. Two variants ($I$ and $II$) of the process of longitudinal US impregnation at different frequencies of the US $f$ of the low-frequency range were investigated. Variant $I$ is FF impregnation with the original PB, while the US tool contacts the surface of the impregnated FF. Variant $II$ is presonicating of liquid PB before impregnation, then impregnation of FF with sonicated PB, while the US instrument is in contact with the surface of the impregnated FF.

The kinetic curves of the process of longitudinal impregnation of glass fiber with liquid oligomer ED-20 at an impregnation temperature of 50°C for the two above-mentioned impregnation variants ($I$ and $II$) are shown in Figure 6.

Thus, when processing the effective part of the US activation of the resin part and the impregnated FF, the wetting angle on the glass substrate $\theta$ decreased by 15–25% (see Figure 5), and the lifting height on the glass fiber $h$ increased in 1.5-3 times (see Figure 6). This clearly indicates an increase in the wetting ability of the EB, as well as an increase in the efficiency (productivity) on average 2–3 times and quality of the FF longitudinal impregnation process under the influence of US.

Thus, the effective US activation of the surface of a continuous FF in the presence of an EC layer improves both the process of preliminary longitudinal impregnation and the process of further dosing of the PB deposition rate, since it prevents fiber breaks, fluffing, and injury.

Such activation also contributes to the degassing (~5–10%) of EC, reducing on average 1.5–2 times internal friction in the structure of a long fibrous material and its straightening before impregnation.

In addition, US activation contributes to the removal (mechanical destruction) of hydrophobic finishing agents, dressing (sizing), and other components from the FF surface, which are applied to the surface of the continuous FF in order to reduce its injury and for better preservation during prolonged storage. Such finishing agents (chemical modification), entering into chemical interaction with both glass fiber and binder, increase (~10–20%) moisture resistance and thermal stability and improve dielectric properties and other characteristics of the material but at the same time reduce (~10–15%) their absorbency due to hydrophobicity.

## 2.7. Physicomechanical Properties of Sonicated Epoxy Matrices.

Experiments on measuring adhesive strength depending on the time $\tau$ of US treatment under normal separation to steel 45 as well as short-term static tensile strength and bending of hardened polymers based on ED-20+DETA showed the following. The change in the adhesive bond strength of a polymer as a function of the sonicating time $\tau$ (see Figure 7) with a normal adhesive strength $\sigma_{na}$ and a shear adhesive strength $\sigma_{sa}$ to steel 45 has an extreme character with a maximum in the region $\tau = 25 – 35$ minutes.

In this case, the greatest hardening at a frequency of $f = 22$ kHz is 37–43%. With an increase in the sonicating time $\tau > 35$ min, the normal adhesive strength $\sigma_{na}$ and shear adhesive strength $\sigma_{sa}$ decrease, and for the sonicating time $\tau = 80$ min, the strength value is even less than the initial value (not shown in Figure 7).

A similar character of changes in the tensile strength $\sigma_{ts}$ and compression strength $\sigma_{cs}$ depending on the sonicating time $\tau$ takes place for the ED-20+DETA polymers (see Figure 8). However, the optimal time interval for sonicating time $\tau$ is shifted to the right and is $\tau = 30 – 40$ min; the amount of hardening is 44–55% for tensile strength $\sigma_{ts}$ and 44–47% for compression strength $\sigma_{cs}$ (see Figure 8).

The obtained experimental results are the basis for the design of the technological process of capillary impregnation of fibrous FF with liquid PB under the action of low-frequency US [44–46]. In particular, in work [44], the results of hardening of sonicated epoxy polymers, which are used as a matrix for impregnation of reinforced fibrous fillers, are investigated. It also analyzes the qualitative and quantitative results of changes in the structural parameters of hardened epoxides, previously treated with US, with various regime parameters.

In study [45], it was shown that the physicomechanical parameters of the hardening of unidirectionally reinforced traditional composites, processed under effective US treatment conditions, reach at least 11–18%. And most importantly, the coefficient of variation of these characteristics is reduced from 16–18% for the original (untreated with US) reinforced composites to 4–5% after such processing.

The developed method of selecting effective technological parameters of US treatment of liquid polymer media and design parameters of US equipment for sonicating liquid rectoplastic materials is analyzed in study [46]. The practical use of the developed technique contributes to the improvement of the technological properties of liquid epoxy matrices, as well as to hardening of traditional reinforced PCMs based on them while ensuring the stability on average 3–4.5 times of the performance characteristics of these PCMs. Also, the obtained results testify to the achievement of energy saving on average 2–2.5 times during US molding of such PCMs as a result of the reduction of the cumulative curing time of the PBs of both "cold" and "hot" hardening [47].

As a rule, effective parameters of the process of US treatment (the main ones are processing time $\tau$, amplitude $A$, frequency $f$, intensity $I$, temperature $T$, and pressure $P$) are determined experimentally in each specific case, for example, using experimental and statistical modeling methods, as well as structural-parametric modeling. In some cases, on their basis, appropriate universal engineering techniques are developed, which are used for other cases, for example, for designing thermoplastic molding processes [48].

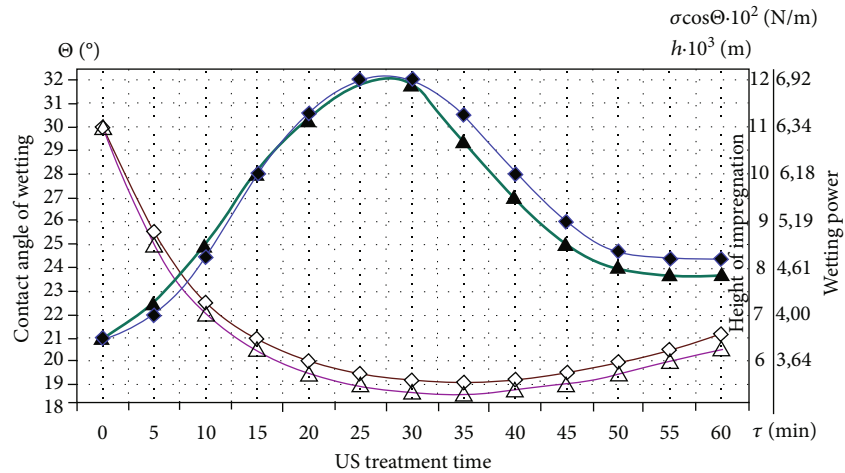

FIGURE 5: Change of wetting angle Θ, wetting ability $\sigma \cos \theta$, and lift height $h$ on glass fiber for EO brand ED-20 at 20°C depending on sonicating time $\tau$ at US frequencies $f$ 16 kHz and 20 kHz: wetting angle Θ (empty rhombus) and wetting ability $\sigma \cos \theta$ (filled diamond) at US frequency $f = 16$ kHz; wetting angle Θ (empty diamond) and wetting ability $\sigma \cos \theta$ (filled diamond) at US frequency $f = 20$ kHz.

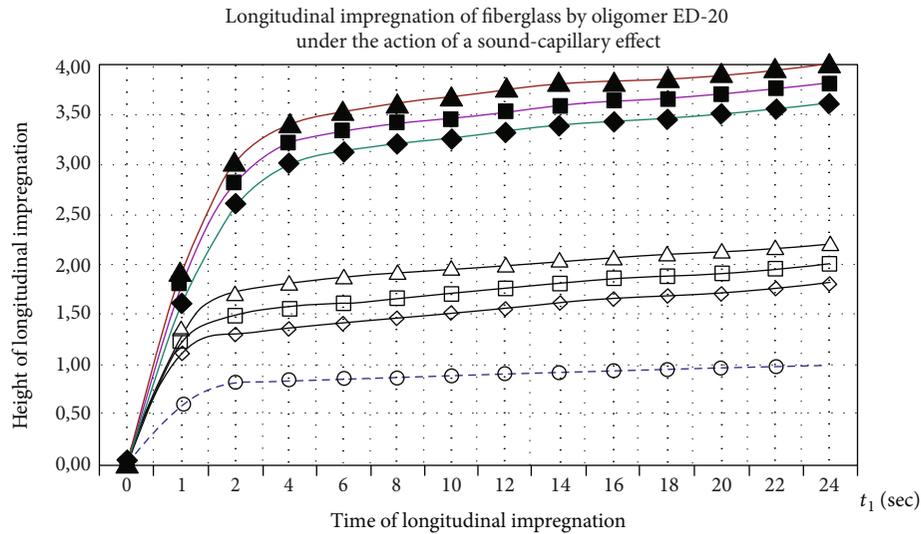

FIGURE 6: Kinetic curves of the process of longitudinal impregnation of fiberglass with an oligomer of ED-20 in two variants (*I* and *II*) of US impregnation: open circle: kinetic curve of the process of "free" capillary impregnation of fiberglass with the original PB; open square: theoretical curve of the kinetic equation of the process of longitudinal impregnation, given in [29], for the first (I) variant of US impregnation, US frequency $f = 18$ kHz; open rectangle: experimental kinetic curve of the process of longitudinal impregnation for the first (*I*) variant of US impregnation, US frequency $f = 18$ kHz; open diamond: experimental kinetic curve of the process of longitudinal impregnation for the first (*I*) variant of US impregnation, US frequency $f = 22$ кГц; open square: experimental kinetic curve of the process of longitudinal impregnation for the first (*I*) variant of US impregnation, US frequency $f = 18$ kHz; filled square: theoretical curve of the kinetic equation of the process of longitudinal impregnation, given in [29], for the second (*II*) variant of US impregnation, US frequency $f = 18$ kHz; filled triangle: experimental kinetic curve of the process of longitudinal impregnation for the second (*II*) variant of US impregnation, US frequency $f = 18$ kHz; filled diamond: experimental kinetic curve of the process of longitudinal impregnation for the second (*II*) variant of US impregnation, US frequency $f = 22$ kHz.

For examples of the developed approaches to the design of high-performance US equipment in the form of US concentrators used in the sonication of liquid PB and impregnation, as well as dosed devices with a rectangular radiating plate, you can refer in studies [49–53]. In the automated design and operation of a vibrational system consisting of an emitting plate with vibrators, it is necessary to perform a number of conditions that provide a resonant mode of operation of these oscillatory systems. However, the main difficulty in designing oscillation systems that contain a rectangular radiating plate is that the sizes of these plates, as a rule, are picked up experimentally, which significantly increases the cost of their manufacturing.

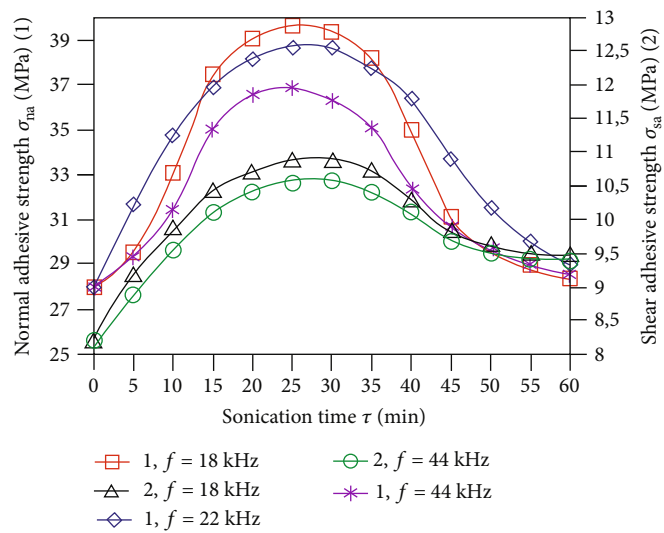

FIGURE 7: The change of normal adhesive strength $\sigma_{na}$ and shear adhesive strength $\sigma_{sa}$ to steel 45 of ED-20+DETA polymer depending on the sonicating time $\tau$ at US frequencies $f = 18$ kHz (open circle), 22 kHz (open diamond), and 44 kHz (open square).

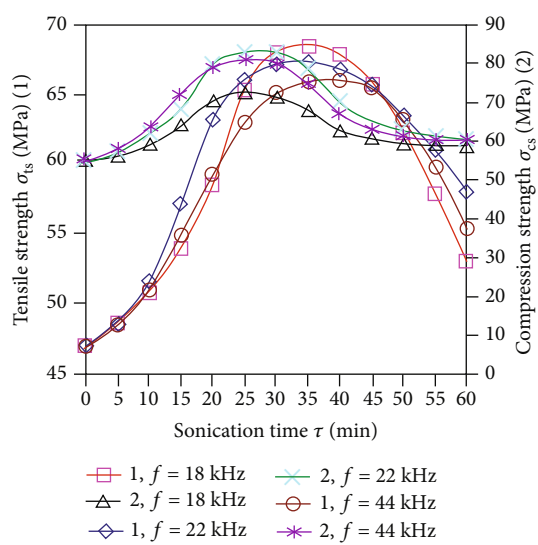

FIGURE 8: The change in the short-term static tensile strength $\sigma_{ts}$ and compression strength $\sigma_{cs}$ depending on the sonicating time $\tau$ at US frequencies $f = 18$ kHz (open circle), 22 kHz (open diamond), and 44 kHz (open square).

Study [49] describes the developed advanced method for the deterministic selection of effective design and technological parameters of equipment that is used for the US modification of liquid polymer compositions and the impregnation of oriented fibrous fillers.

The use of the technique allows to increase the productivity of US processing, including the acceleration of the processes of impregnation and dosed deposition, during the molding of traditional thermoplastic PCMs. Features of the calculation of US cavitators based on piezoceramic transducers attached to the base of the radiating plate, which is sub-jected to bending longitudinal and transverse oscillations, are described in investigation [50].

Analytical acoustic dimensions of the components of a sectional piezoelectric transducer, which is used in the process of contact US treatment of dry tissue 1120 mm wide, which is soaked with liquid PB, are proposed. Effective technical means of US cavitation treatment and dosed deposition of thermoplastic binders on impregnated fibrous fillers are discussed in work [51]. A technique is proposed for designing the parameters of US cavitators with rectangular radiating plates as applied to a specific technological process. The technique of computer-aided design of US cavitation devices with rectangular emitting plates and its parametric visualization are presented in [52]. Parametric models with elements of computer graphics graphically reflect the existing variational dependencies between the elements of the designed US cavitation devices.

In the continuation of the approaches described in the studies [51, 52], the article [53] discusses the main techniques of computerized three-dimensional design of an US metering device with a rectangular radiating plate based on the developed methodology of structural-parametric geometric modeling. In the framework of this methodology, the use of the system approach, which involves the representation of any object or the process of its formation as a certain ordered set of some individual components of the components, which, in turn, may include other elements, is preferable. With the practical use of this methodology, the principles of integrated approach, variability, and optimality, as well as the principle of openness and development, are widely applied, which envisages the possibility of easy updating and expansion of components. The proposed approach allows you to optimize the parameters of the above metering device.

Another application of the methods of physicochemical modification of polymers on the example of developing efficient equipment and technologies for connecting and repairing polyethylene (PE) pipelines using US modification and heat shrinkage is considered in studies [54–58]. Moreover, the article [54] considers the efficiency of using epoxy adhesive compositions and the method of their US modification. The above means are used in repair technologies for connecting and restoring polymer pipelines intended for the transportation of low and medium pressure gas. A positive result of the application of the above means was noted, which is to reduce the level of residual stresses in the pipelines (~12–15%) and to increase their service life on average 1.5 times.

The use of unmodified shrink couplings (couplings with shape memory effect) and adhesive compositions formed from epoxy polymers when connecting and restoring polymer pipelines for transporting gas is analyzed in [55]. In particular, the improvement of the deformation-strength and adhesive characteristics of the obtained epoxy composites was noted; for example, the maximum deformation of the tubular blanks of the studied epoxy polymers, which did not lead to their destruction, was at least 15–17%. Aspects of molding epoxy repair couplings with shape memory, obtained on the basis of US modification of components of epoxides, are described in study [56]. The change in the strength of the heat shrink couplings depending on the parameters of their US treatment is analyzed, and effective operating parameters of the US treatment are found under which the maximum strain hardening of the heat shrink couplings is simultaneously achieved ensuring an acceptable degree of deformation ($\varepsilon = 5 - 8\%$) resulting from the heat treatment of the couplings.

In particular, the following results were obtained: relaxation of residual internal stresses observed in tensile products to 70%, increase in mechanical characteristics up to 40%, increase in operational resource (durability) of products by 5–8 times, and increase in productivity of the heat treatment process in comparison with traditional technologies of quenching and annealing by 5–10 times.

The effectiveness of complex US effects in the preparation of epoxy adhesive compositions, which are used in the impregnation of glass tape in bandage technologies for the repair of PE pipelines, as well as a range of methods for modifying the surface of connected PE pipes, is discussed in [57]. The developed methodology for the implementation of the production bases of the developed banding technology is described.

Using the developed technology allows to obtain the following qualitative results: it provides almost complete penetration of the binder (adhesive mixture) into the impregnated and wound fiberglass tape filler and also helps to remove air inclusions that inevitably form during winding, especially in winter.

The hardware for the process of thermistor welding during the repair of low and medium pressure PE pipelines, as another method for modifying the surface of PE pipelines using temperature, was described in work [58]. It is noted that the found effective parameters of the process of thermistor welding vary depending on the diameter of the connected PE pipelines, as well as the sizes of the used thermistor couplings.

The method described in [58] for repairing PE pipes using thermistor welding and thermistor couplings is used in the construction and repair of underground gas pipelines of low ($P = 0.05$ MPa) and medium ($P$ up to $0.3$ MPa) pressures made of PE pipes. It was experimentally established that such welded joints are more reliable (on average 1.5–2 times) compared to the standardly used means for repairing PE pipes due to better contractual interaction between the inner surface of the heated thermistor sleeve and the outer surface of the PE pipe.

The above approaches of using effective means of US processing determine the possibility of improving the surface properties of FFs, improving the technological and operational characteristics of PBs, and improving the performance of the molding process of such PCMs, as well as significant energy savings (~30–50%) during their molding.

### 2.8. Ultrasonic Modification during Obtaining Nanocomposites.

In the synthesis of new NM PCMs of deterministic functionality, such an accompanying phenomenon that occurs at high US intensities such as sonochemistry is often used [59–63].

The subject of sonochemistry is the study of the use of high-intensity US in chemical reactions and processes. At the same time, at low US intensities, dispersion of nanofillers in liquid, including polymeric, media is traditionally carried out, obtaining dispersions and nanosuspensions, as well as products derived from them [64].

It is also necessary to note the widespread use of universal epoxy matrices for the synthesis of NM PCMs of a new class with new nanofillers such as graphene oxide, graphene aerogel, graphene foam, and other derivatives [65–72]. The optimal incorporation of nanofillers, in particular, carbon ones, improves the operational and technological properties of NM PCMs, for example, physical and mechanical properties, chemical resistance, electrical conductivity, and thermal conductivity [73].

For example, the use of PBs modified with CNTs in hybrid PCMs makes it possible to increase the crack resistance coefficients $G_{1c}$ and $G_{2c}$ by an average of 15–40%. So, the following physicomechanical and operational characteristics were obtained for NM PCMs based on 1750 stretched sheets from CNT array oriented in mutually perpendicular directions: tensile strength 117 MPa, elasticity modulus 7.45 GPa, and electrical conductivity 2205 S/cm at a CNT concentration of 8%.

Polydimethylsiloxane rubbers (polysiloxanes) are known to possess such valuable properties as gas permeability, thermal and chemical resistance, high dielectric properties, biological inertness, and compatibility with a living organism [74]. Therefore, when dispersing nanoparticles in such liquid media to increase the chemical activity of the surface of carbon nanomaterials [74], two main methods (stages) are used: the single-stage method and the two-stage method [75].

The first stage consists in dispersing nanoparticles in a base fluid (solvent, etc.) and in obtaining a suspension. In the second stage, the resulting suspension is mechanically treated (including US) to reduce the aggregation (agglomeration) of the nanoparticles. The results of work [75] show that the thermal conductivity of the fluid-particle system depends on both the particle volume fraction and the shape.

Assuming that the sphericity of copper nanoparticles is 0.3, the thermal conductivity of water can be enhanced by a factor of 1.5 at the low volume fraction of 5% and by a factor of almost 3.5 at the high volume fraction of 20%. This finding demonstrates that, theoretically, the feasibility of the concept of nanofluids, i.e., metallic nanoparticles, are capable of significantly increasing the thermal conductivity of conventional heat transfer fluids. In particular, it was noted that one of the benefits of nanofluids is dramatic reductions in heat exchanger pumping power. For example, to improve the heat transfer by a factor of 2, the pumping power with conventional fluids should be increased by a factor of ~10.

With regard to the innovative nanofiller in the form of graphene, it should be noted that there is a dependence of the degree of dispersion of graphene (and graphene oxide) on the completeness of the use of the potential of graphene, which has truly unique operational properties [76].

According to existing estimates, graphene has high mechanical rigidity and records high thermal conductivity ($\sim 1\,TPa$ and $\sim 5 \cdot 10^3 W \cdot m^{-1} \cdot K^{-1}$, respectively). The high mobility of charge carriers (maximum electron mobility among all known materials) makes it a promising material for use in a wide variety of applications. So, for example, graphene is known to conduct a thousand times more current than copper conductors, and ten times less energy is released in the form of heat. In this case, the properties of graphene vary depending on its structural parameters [77]. And the use of US treatment contributes to the weakening of interlayer forces and allows deterministic control of important processing parameters during the production of graphene and its derivatives.

This is confirmed by the results of the review [78], which describes the possibility of obtaining small and even colloidal suspensions as a result of US processing of a graphene dispersion, characterized by a high concentration of 1 mg/ml and the possibility of subsequent production of relatively clean graphene sheets with a conductivity of 712 S/m.

The effect of US on the thermal conductivity and viscosity of CNT suspensions (0.5 wt.p.) in an ethylene glycol solvent-based nanopowder was studied in investigation [79]. In this case, gum arabic with a concentration of 0.25 wt.p. was used as a dispersant for MWCNTs.

It is known that graphene is structurally a flat grid of carbon atoms. The latter is located at the corners of regular hexagons at a distance of 0.1418 nm [80].

The efficiency of using graphene as an additive in NM PCMs is due to the fact that this improves the electrical, physicomechanical, and barrier properties of the nanocomposite at very low loads [81]. A convenient sonochemical method was described for the preparation of polystyrene functionalized graphenes starting from graphite flakes and a reactive monomer, styrene. US irradiation of graphite in styrene results in the mechanochemical exfoliation of graphite flakes to single-layer and few-layer graphene sheets combined with functionalization of the graphene with polystyrene chains. The polystyrene chains are formed from sonochemically initiated radical polymerization of styrene and can make up to ≈18 wt.p. of the functionalized graphene, as determined by thermal gravimetric analysis. This one-step protocol can be generally applied to the functionalization of graphenes with other vinyl monomers for graphene-based composite materials. A photograph of a graphene sample in tetrahydrofuran after 6 months at room temperature showed that the sample was without precipitation.

It was noted that low-frequency US is a potentially acceptable method of producing high-quality graphene and on an industrial scale [78, 82]. It was developed several ways to use US when receiving graphene. A reduced graphene oxide- (RGO-) gold (Au) nanoparticle (NP) nanocomposite was synthesized by simultaneously reducing the Au ions and depositing Au NPs onto the surface of the RGOsRGO simultaneously. To facilitate the reduction of Au ions and the generation of oxygen functionalities for anchoring the Au NPs on the RGOsRGO, US irradiation was applied to the mixture of reactants. As a result, the dense and uniform deposition of nanometer-sized Au NPs was observed on the RGO sheets from the TEM images. Besides, US application can readily take Au-binding-peptide-modified biomolecules, readily implying its possibility in possible biological applications.

Despite this, US graphene production is represented by a simple one-step process [83]. First, graphite is added to a mixture of dilute organic acid, alcohol, and water. After that, the resulting combination is subjected to sonication. Organic acid acts as a "molecular wedge," separating the graphene sheets from the "parent" graphite. The principal possibility of obtaining ultracapacitors with unique characteristics, namely, specific capacity (~120 F/g), specific power (~105 kW/kg), and energy density (~9.2 W/kg), has been declared. This technology eliminates the use of strong oxidizing agents or reducing agents.

The preparation of graphene by exfoliation under the influence of US was described in study [84]. Reduction of a colloidal suspension of exfoliated graphene oxide sheets in water with hydrazine hydrate results in their aggregation and subsequent formation of a high-surface area carbon material which consists of thin graphene-based sheets. Images of graphene oxide, detached as a result of US treatment at concentrations of 1 mg/ml in water, indicated the presence of sheets with an almost uniform thickness of ~1 nm.

The study [85] describes the process of obtaining sheets of pure graphene in the production of nonstoichiometric graphene nanocomposite $TiO_2$. At the same time, pure graphene nanosheets were made from natural graphite using an US reactor, in which a static pressure of 0.5 MPa was maintained. The concentration of graphene as a dopant varied from 0.001% to 1%. The preparation of graphene nanoribbons in two stages was described in [86]. At the first stage, the graphene layers of graphite were weakened due to heat treatment at a temperature of 1000°C for one minute in 3% hydrogen in argon gas. At the second stage, graphene was decomposed into strips using a low-frequency US. The resulting nanoribbons are characterized by much more "smooth" edges compared to nanoribbons produced by conventional lithographic means.

US treatment also facilitates the twisting of graphene monolayers in an aqueous medium, for example, in a hydrogel [87]. It was found that US treatment helps the scrolling up graphene monolayers into carbon nanoscrolls. As a result, a high efficiency (~80%) of the conversion of monolayers into nanoscrolls was achieved, which makes the production of nanoscrolls promising for commercial applications.

The possibility of choosing the form of a graphene hydrogel by choosing the desired shape of the US is described in [88]. The uniaxial compression tests of the graphene hydrogel at different set strains were made. The curve shows an initial linear region at deformation $\varepsilon < 10$ %, and a plateau with gradually increasing slope until very high strains up to 60%. The elasticity modulus $E$ and yield stress of the graphene hydrogel were measured to be about 0.13 MPa and 28 kPa, respectively.

The impact of low-frequency US is also effective for the formation of a three-dimensional structure in the synthesis of graphene-based hydrogels. This is confirmed by the results of work [89], in which the effect of low-frequency US treatment on the formation of a graphene hydrogel was investigated. It is noted that the hydrogels obtained using an effective US treatment are characterized by low critical concentrations of gelation in the range from 0.050 mg/ml to 0.125 mg/ml.

Thus, the considered direction of designing the parameters of the process of US modification of the components of composites, including FF and a liquid polymer matrix, when producing traditional PCMs and synthesizing NM PCMs, has the prospect of further development due to the high efficiency of US exposure compared to other well-known modification methods.

The possibilities of using US in melting polymer melt in the channels of a screw extruder, as well as the corresponding physical and mathematical models of the process of melting polymer melt, were proposed in study [90]. The developed approach allows predicting the efficiency of incorporating nanoparticles into a highly viscous thermoplastic matrix, as well as improving the homogenization of the resulting mixture. Also, using an example of polystyrene research, it was found that even a small (within tenths of a percent) number of nanoparticles in the polymer matrix increase the thermal stability of the nanomodified material by about 50°C. Moreover, a further increase in the content of nanoparticles in the thermoplastic polymer does not significantly affect its thermal stability.

The directions of creating functional-enhanced NM PCMs, among which, in particular, are improving the surface properties of nanomodifiers used, improving their deagglomeration processes and subsequent homogenization in a liquid medium, modifying the polymer matrix, and developing innovative methods for synthesizing structural hybrid carbon plastics of combined filling, are discussed in a survey study [91]. The efficiency of creating carbon fiber PCMs, as well as the prospects for creating these materials based on reinforcing fabric with NM fillers, is described. The methods of obtaining functional NM carbon-composites with improved physicomechanical and operational properties, in particular, with increased strength and electrical conductivity, are shortly characterized.

The current state of the art of the US treatment technology applied to polymer melts is presented in review [92]. It was mentioned that for nearly 50 years of innovation, the US treatment technology of polymer melts has demonstrated a great versatility as seen in many applications such as improvement in the processability of polymers, devulcanization of rubbers, preparation of polymer blends, and recently preparation of NM PCMs. But, according to the authors [92], the fundamental mechanisms responsible for these effects remain enigmatic, thus limiting the fullest exploitation of the US technology for upcoming polymer melt processing.

### 2.9. Some Methods of Physicochemical and Biological Modifications of Components of Polymers of Functional Application.
The study of surface interaction of substrates in polymer composites is an important area of research not only in nanotechnology but also in biotechnology and in a number of other fields. In addition, the science of the surface interaction of components is important in the development, manufacture, and application of various kinds of coatings, including polymeric. This is due to the fact that the surface properties of the polymer matrix and the substrate material, as well as the resulting adhesion and cohesion forces, largely determine the reliability and functional applicability of the designed composite.

It should be noted that the surface properties of modern materials are usually uninformative from the point of view of ensuring acceptable wettability, adhesive properties, biocompatibility, etc., which predetermines their functional application. Therefore, it is desirable to modify the surface of these materials before use or any further processing, such as, the application of deterministic coating with functional materials. Consequently, to obtain the desired surface quality of materials, their morphological properties, chemical structure, and composition should be deterministically changed.

Thus, when implementing a chemical modification, a polymeric material can be functionalized, for example, by adding small fragments, as well as oligomers and even other polymers (grafting copolymers), to the interface or to the "matrix-material" interface. As for nanofibers, their specific application is determined by the surface properties of the latter. In particular, the surface nanofiber properties affect the wettability, electrical conductivity, optical properties, and biological compatibility of the filled composite material.

Physical modification methods, in addition to the previously mentioned methods, also include such a widely used method as surface oxidation. Such a modification is mainly represented by the methods of plasma treatment, corona treatment, and flame treatment. The mechanism of action of the above methods includes the splitting of polymer chains in a polymer material and the subsequent incorporating in the "work" of carbonyl and hydroxyl functional groups.

In turn, the incorporating of oxygen into the surface during oxidation contributes to the creation of a higher surface energy, which makes it possible to more fully and qualitatively coat the substrate.

The advantage of plasma treatment in comparison with other physicochemical modification methods is that, as a result of this treatment, higher values of the interfacial energy are achieved, and more intensive injection of monomer fragments is observed. At the same time, the plasma is thermodynamically unfavorable, and besides, limited flows prevent the achievement of high plasma processing rates. As a result, plasma-treated surfaces are usually characterized by heterogeneity, variation in consistency, and inconsistency in characteristics. All of these prevent the use of plasma processing as a competitive method for modifying the surface of components in an industrial scale.

The latest achievements in the field of surface functionalization of polymeric materials using gas plasma treatment are presented in [93]. Attention is drawn to the fact that the functionalization of such materials is due to the chemical interaction between solid materials and active plasma components. Among the active components of the plasma emit charged particles, neutral radicals, excited particles, and UV radiation. It is indicated that the degree of functionalization of the surface of the material being processed, as a rule, depends on the type of polymers and the characteristics of the flows of the active forms of the plasma.

Therefore, to determine the parameters of plasma treatment, it is possible to change the wettability of the surface of the materials being processed, that is, to regulate the adhesion. At the same time, there is a difference in the characteristics of surface properties obtained by different authors for the same materials subjected to plasma treatment. The reason for this, apparently, is the difference in the parameters of the gas discharges used to generate the plasma. It was especially noted that numerous authors have used plasma treatment for modification of the surface functionalities. However, the obtained surface properties often differ even for the same materials, which does not allow recommending specific plasma treatment parameters for usage.

Compared to plasma treatment, flame treatment is a controlled, fast, and cost-effective method for increasing surface energy and improving the wettability of modifiable composite components, such as polyolefins and metal substrates. With such a high-temperature plasma treatment, ionized gaseous oxygen is used, which is incorporated over the surface by means of a jet flame. As a result, polar functional groups are added during the melting of surface molecules, and subsequently, they are fixed at the processing site when the material is cooled.

In a survey study [94], aspects of the functionalization of the surface of polymer nanofibers by various methods are considered. In particular, the results of ion-beam implantation with surface modification of polymer nanofibers were described. It was stated that the elastic modulus of the fiber was increased by 30% at a dose of $10^{15}$ ions/cm$^2$ of the nitrogen ion. It was also concluded that the dominant chemical modifications by treatment with nitrogen ions are the incorporation of amine and amide functional groups, which are used to ensure compatibility of cells on polymer surfaces.

The main goal of work [95] was to improve the methods of solid-phase technology for processing complex polymeric materials, as well as NM PCMs using the electrophysical effects of microwave and US fields. The authors in [95] declare that effective processing regimes contribute to improving equipment performance, achieving energy saving, and improving the performance characteristics of the materials obtained. In particular, the effectiveness of ultrasonic treatment is to improve the physicomechanical properties, namely, the tensile failure stress, as well as the modulus of elasticity by 20-50%, the failure stress in shear by 30-35%, and the reduction of the required forming pressure by 1.5-2.5 times.

Recent trends in nanostructuring and functionalization of solid materials to improve their functional properties are analyzed in study [96]. The main focus of the authors is on the study of the morphology of the surface of solid materials. This is due to the fact that morphology predetermines the effective surface area, which substantially exceeds the macroscopic geometric surface area of solid materials. It is also noted that most of the technologies used in mass production of PCMs, including casting, injection molding, extrusion, and rolling, provide limited opportunities for increasing the contact surface area of solid materials that exceeds the geometric surface area.

In some cases, solid materials are obtained from liquids whose surface energy, due to the physical nature, a priori contributes to the formation of smooth surfaces. Moreover, even after hardening, such a smooth surface is often preserved. Despite this, a solid material with a large surface area (and much larger than the geometric surface area) is still preferable. The author in [96] notes that the rough surface often provides better adhesion of the polymer coating to the substrate and is therefore preferred in many cases. For example, the contact surface area may be significantly increased compared to the geometric surface area by etching or deposition. And increasing the plasma power while keeping the precursor flow rate constant results in faster deposition rates and subsequently thicker coatings, but with a higher oxide content. All conditions studied in [96] lead to superhydrophobic surfaces with static contact angles higher than 150° and contact angle hysteresis lower than 6°.

Separately, it is necessary to dwell on the aspects of the use of composite materials in biological and medical applications. Currently, polymers are widely used in biomedicine because of their biocompatibility and good mechanical properties, which in some cases approach the mechanical properties of human tissues. At the same time, these materials are, as a rule, chemically and biologically inert, which determines the need for their modification for successful use. Therefore, the functionalization of polymer surfaces is the basis for the production of new biologically compatible materials.

Surface characteristics of biocompatible polymers, such as topography (on a macro-, micro-, and nanoscale), features of surface chemistry, surface energy, charge, or wettability, are interrelated. These characteristics together predetermine the biological characteristics of materials used for biomedical applications. In particular, they regulate the biological response at the interface of the implant/tissue (for example, affect the adhesion and orientation of cells, their mobility, etc.).

The dominant factors governing the use of biological polymeric materials are the body's response to extraneous material, that is, biological compatibility, which, in turn, depends on the surface interactions of the material used and the liquid matrix. Therefore, in addition to physical, chemical, and physicochemical modification methods, there is also a biological modification used to create materials for biomedical purposes [97].

For example, the surface of a biological material is often modified using mechanisms activated by light radiation (such as phototransplantation). This allows functionalization of the surface of the modified material without damage to its bulk mechanical properties. Surface modification to preserve biologically inert polymers has found wide application in biomedical applications such as cardiovascular stents and in many skeletal prostheses. Effectively functionalized polymer surfaces are able to inhibit protein adsorption, which could otherwise initiate cellular destruction on the implant. The latter is the predominant mechanism for the destruction of medical prostheses.

One of the current trends in the study of cell biology, regenerative medicine, tissue engineering, and many other biological applications is the development of a polymer surface characterized by rational design. It uses chemical, topographic, and mechanical cues to control the behavior of cells. So, the result of tissue engineering studies using scaffolds formed using fiber-optic technologies [97] is material poly(L-lactic acid). It is used for the tendon and ligament (human mesenchymal stem cells). Its mechanical properties are tensile strength: 6.57–7.62 MPa and Young's modulus 47.6–55.0 MPa.

The results of a study [98] showed that physicochemical plasma treatment in an Ar+ medium contributes to the biological functionalization of polydimethylsiloxane. This is a multistep procedure that is completed by immobilizing collagen using a flexible compound of polyethylene glycol of an average length of 6000 Da. The study [98] also states that microarrays on the surface of biological material create structures of different sizes and shapes that control the spread. However, there was no direct correlation between hydrophilicity and initial cellular interaction, probably due to the simultaneous influence of other factors such as superficial chemical structure and topography.

The review [99] analyzes various types of polymer surfaces designed to control the behavior of cells. A detailed discussion of the basic mechanisms is given, and a sufficient number of examples of biomimetic research is given. So, f.e., it was stated that surface energy is an important chemical factor on polymer surfaces of cells. In particular, it can be considered as a measure of the energy of unsaturated bonds that arise as a result of dangling bonds of a surface material. The surface free energy of the modified poly-3-hydroxybutyrate (P3HB) membranes importantly increased after plasma treatment $53.5 \pm 0.9$ mN/m and $57.4 \pm 0.9$ mN/m, respectively, for oxygen-modified and ammonia-modified P3HB. However, only ammonia plasma-treated P3HB membranes exhibited a higher degree of fibroblast cell adhesion and proliferation when compared to both the untreated and oxygen-treated P3HB. This is apparently due to the fact that the polar component of the ammonia plasma-treated P3HB is higher (27.5 mN/m) than that of the oxygen plasma-treated P3HB (18.5 mN/m).

Despite their capabilities and sufficient knowledge, the above methods of physicochemical modification with respect to biomedical polymers have a number of limitations that prevent their widespread use. In this regard, laser methods, in particular, laser texturing of the surface, can be one of the alternatives to the above-mentioned methods of physicochemical modification.

In particular, in study [100], a general review of the scientific and practical foundations and results of applications of laser surface texturing for surface modification of polymers, which are currently used in clinical practice, are presented. Experiments were conducted using a fs Ti:Sapphire laser source ($\lambda = 800$ nm and pulse duration of 150 fs) to evaluate the wettability changes on poly(methyl methacrylate) (PMMA), which is used for fabrication microchannels with controlled size and roughness for microfluidic applications, after the laser treatment. Thus, after the laser treatment, the contact angle reduced from 74.1° (for the base material) up to 20° approximately.

Thus, the spectrum of currently used physical, chemical, physicochemical, and biological methods for modifying the components of polymeric materials tends to be constantly developed and diversified in the direction of improving the surface properties of the materials used for deterministic functional applications.

## 3. Conclusions

The article analyzes the possible directions of improving the operational properties of traditional PCMs and NM PCMs for various functional purposes according to the approach under consideration, which provides for three main research units: improving the surface properties of oriented, dispersed, and nanoscale fillers of traditional and NM PCMs, improving the properties of the liquid polymer matrix and the composite as a whole, and improving the structure of the composite and optimizing the ratio of its components. Moreover, the methods of physical, chemical, and combined physicochemical modification are the main external factors contributing to the declared improvements in the operational properties of filled reactoplastic PCMs by improving the surface properties of the fillers and the technological properties of the liquid polymer matrix.

An epoxy matrix is considered as a polymer matrix due to a complex of valuable operational properties manifested in the composite, as well as good knowledge and industrial development. Widely used classes of fillers, such as oriented macrofillers, dispersed fillers, and nanoscale fillers, are considered as the main types of fillers for modern composites.

The mechanism of interaction of dispersed fillers with a liquid polymer matrix is considered in the framework of the study clustering and physicochemical modification of epoxy polymers with disperse fillers, as well as an energy approach. In particular, it is shown that the structure of the boundary polymer layer is formed as a result of the tendency of dispersed filler particles to reduce their surface energy, and it is energetically more advantageous when the limiting layers of individually dispersed particles begin to interact with each other in contact. As a result of effective chemical modification of the surface of the macrofiller and an increase in its wetting ability, the porosity of the structure of the hardened traditional epoxy PCMs can decrease in value, depending on the type of macrofiller used and a polymer matrix.

It is shown that the development of effective modes of low-frequency US cavitation treatment is an effective direction in the implementation of physical and physicochemical modification methods for creating various polymer composites. Besides that, for nearly 50 years of innovation, the US treatment technology of polymer melts has demonstrated a great versatility as seen in many applications such as improvement in the processability of polymers, devulcanization of rubbers, preparation of polymer blends, and recently preparation of NM PCMs.

It is shown that, in the best case, such an US modification is aimed at intensifying (on average 2–3 times) the impregnation process and dosed application for traditional PCM technologies, for degassing a liquid polymer matrix and for deagglomerating nanofillers and subsequent homogenization of a nanofilled polymer matrix during molding or synthesis of a nanocomposite.

Also, the use of US treatment is effective in obtaining functional nanocomposites with improved surface properties, as well as for improving the physicomechanical and operational characteristics of the obtained materials based on them. So, for example, for traditional epoxy PCMs, the amount of hardening is 44–55% for tensile strength and 44–47% for compression strength, respectively. Also, the obtained results testify to the achievement of energy saving on average 2–2.5 times during US molding of such PCMs as a result of the reduction and dosed curing time of the PBs of both "cold" and "hot" hardening.

In addition to the above three modification methods, some features of the biological modification used to create materials for biomedical purposes are briefly discussed. The materials used for these purposes are usually chemically and biologically inert, which determines the need for their modification for successful use. It was mentioned that the surface properties of modern materials are usually uninformative from the point of view of ensuring acceptable wettability, adhesive properties, biocompatibility, etc., which predetermine their functional application. Moreover, such modification methods as surface oxidation (plasma treatment, corona treatment, and flame treatment), microwave, light radiation (UV), and laser methods are widely used.

So, the spectrum of currently used physical, chemical, physicochemical, and biological methods for modifying the components of polymeric materials tends to be constantly developed and diversified in the direction of improving the surface properties of the materials used for deterministic functional applications.

The results of the review can be used to develop new directions for the creation of both traditional and NM reactoplastic and thermoplastic PCMs with improved surface properties and various functional application.

## Conflicts of Interest

The authors declare that there is no conflict of interest regarding the publication of this paper.